\documentclass[a4paper,10pt]{article}
\usepackage[utf8]{inputenc}
\usepackage{xspace}			 
\usepackage[english]{babel}
\usepackage[T1]{fontenc}
\usepackage{lmodern}
\usepackage{authblk}

\usepackage[margin=25mm,includeheadfoot,bindingoffset=10mm]{geometry}
\usepackage{setspace}
\usepackage{ragged2e}
\usepackage[all]{xy}
\usepackage{graphicx} 
\usepackage{color}
\usepackage[svgnames]{xcolor}

\usepackage{pdfpages}
\usepackage{amsmath, mathtools}
\usepackage{amssymb,amsfonts,dsfont}
\usepackage{mathrsfs}
\usepackage{amsthm}
\usepackage{bm}				
\usepackage{bbm,upgreek}
\usepackage{makecell}
\usepackage{cellspace}
\usepackage{hyperref} 
\usepackage{cleveref}
\hypersetup{linktocpage,colorlinks,citecolor={blue},pdfdisplaydoctitle=true,pdfpagemode=UseOutlines,bookmarksnumbered=true}
\usepackage[sort&compress,numbers]{natbib}
\usepackage{url}
\usepackage{doi}
\usepackage[small,bf,singlelinecheck=false]{caption}

\newcommand{\bra}[1]{\langle #1 |} 
\newcommand{\ket}[1]{| #1 \rangle } 
\definecolor{cbl}{rgb}{0,0,1}
 
\definecolor{crd}{rgb}{1,0,0}
 
\newcommand{\upd}{\mathrm{d}}
\newcommand{\tr}{\mathrm{tr}}

\newcommand{\ie}[0]{\textit{i.e.} }
\newcommand{\eg}[0]{\textit{e.g.} }

\newcommand\e{\mathrm{e}}

\title{Continuous collapse models on finite dimensional Hilbert spaces}
\author{Antoine Tilloy\footnote{\url{antoine.tilloy@mpq.mpg.de}} }
\affil{\vskip-0.2cm \it \small Max-Planck-Institut f\"ur Quantenoptik, Hans-Kopfermann-Stra{\ss}e 1, 85748 Garching, Germany \normalsize}
\date{\small (\today) \normalsize}

\begin{document}

\maketitle

\noindent \textit{Chapter prepared for the book ``Do wave functions jump? Perspectives on the work of GC Ghirardi'', editors: V. Allori, A. Bassi, D. D\"urr \& N. Zangh\`i; Springer International Publishing}

\section{Introduction}
Collapse models come in many flavors, with varying levels of complexity. Yet even the simplest physically realistic models have a phenomenology that is non-trivial to study rigorously, if only because continuous space imposes an infinite dimensional Hilbert space. Here, we would like to focus on toy models, that apply to finite dimensional Hilbert spaces, that can be efficiently simulated, and are amenable to a precise and to some extent rigorous study. We shall mostly be interested in collapse models for qubits, \ie with $\mathscr{H}=\mathds{C}^2$, that is for the simplest quantum mechanical system one can think of.

The prototypical equation we will discuss gives the dynamics of a probability (or population) $p_t = |\bra{\psi_t}              +\rangle_z|^2 \in [0,1]$ for a qubit to be in one state (say its ground state, or spin-up state $\ket{+}_z$) which is a fixed point of the collapse process. It is an It\^o stochastic differential equation that reads:
\begin{equation}\label{eq:basic}
\upd p_t = \underset{\text{``regular'' dynamics}}{\underbrace{\lambda \, (p_\text{eq} - p_t)\,\upd t}} + \underset{\text{collapse dynamics}}{\underbrace{ \sqrt{\gamma}\;  p_t (1-p_t) \,\upd W_t}},
\end{equation}
where $p_\text{eq}\in ]0,1[$ is a constant equilibrium probability in the absence of collapse, $W_t$ is a Wiener process (Brownian motion), $\lambda$ is the rate or frequency associated to the dynamics in absence of collapse and $\gamma$ is the rate\footnote{Note that the inverse time scale $\gamma$ appears with a square root, intuitively because $\upd W $ scales like $\sqrt{\upd t}$.} associated to collapse dynamics. We shall explain later how this equation is obtained, but let us briefly give an intuition for its phenomenology.

The dynamics in absence of collapse, if taken alone, yields an exponential convergence (controlled by the rate $\lambda$) to the equilibrium probability $p_\text{eq}$. It is typically the dynamics one obtains by coupling a qubit to a thermal bath. On the other hand, the collapse term induces an inhomogeneous diffusion with a coefficient that vanishes in $p=0$ and $p=1$. Hence, under this dynamics, the probability wanders in an unbiased way until it reaches one of these two fixed points where the dynamics freezes. In brief, thermalization dynamics deterministically drives the probability to $p=p_\text{eq} \in ]0,1[$, while collapse stochastically drives it to $p=0$ or $p=1$. It is from this competition that rich dynamics can emerge.

\begin{figure}
    \centering
    \includegraphics[width=0.8\textwidth]{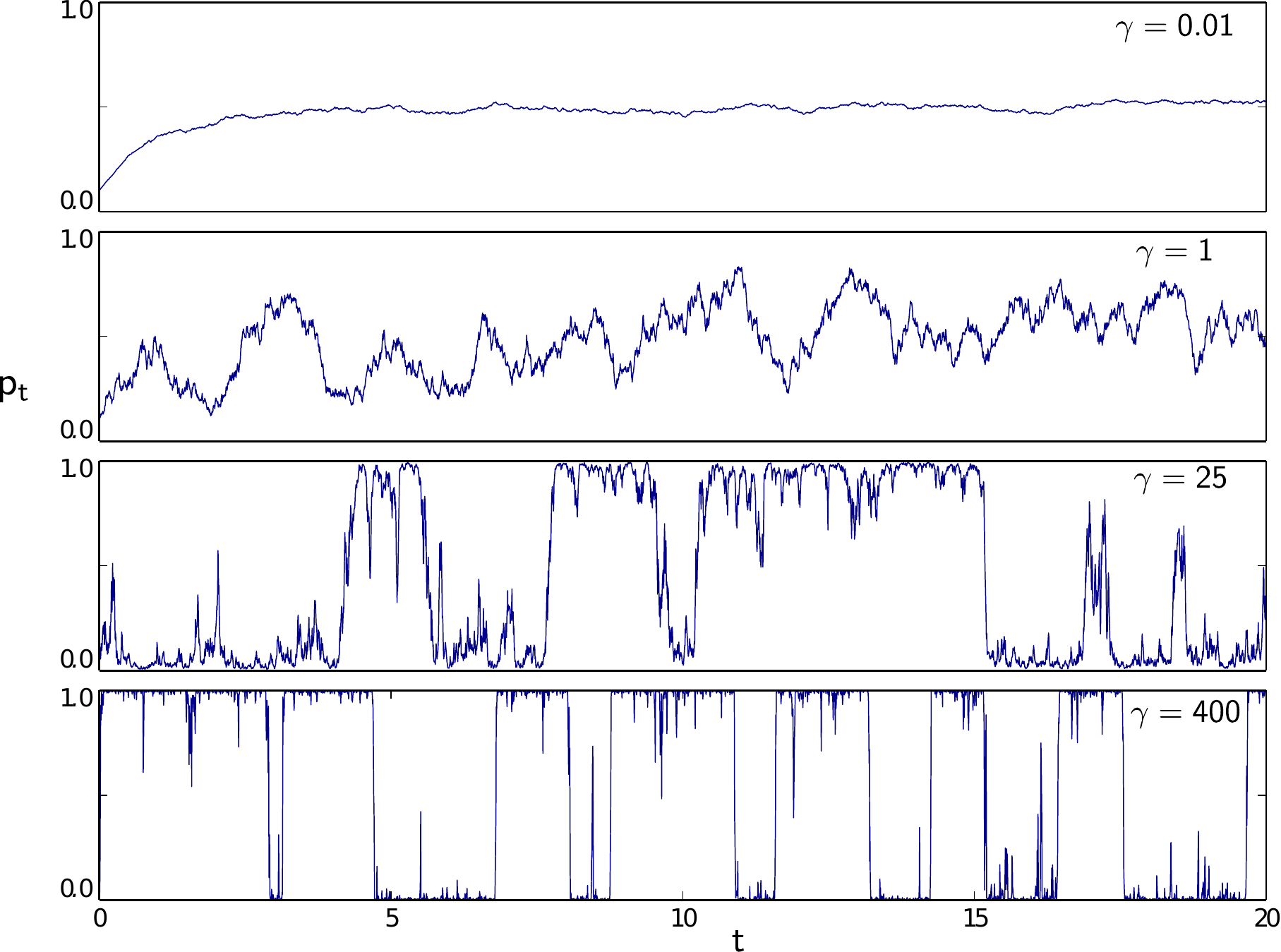}
    \caption{\textbf{Trajectories $p_t$ from } $\upd p_t =\lambda \, (p_\text{eq} - p_t)\,\upd t + \sqrt{\gamma}\,  p_t (1-p_t) \,\upd W_t$ \textbf{ for increasing values of $\gamma$} -- Typical trajectories are shown for $\lambda = 1$, $p_\text{eq}=0.5$, and $\gamma = \{0.01, 1, 25, 400\}$. They are obtained through a naive Euler discretization of \eqref{eq:basic} with $\upd t = 5.10^{-4}$. Far smarter discretization schemes can be used for this particular type of stochastic differential equations (see \eg \cite{rouchon2015}), and they confirm this qualitative behavior. In particular, the sharp, almost punctual excursions decorating the jump process in the last plot are \emph{not} numerical artifacts.}
    \label{fig:summary}
\end{figure}

Before saying more, it is instructive to look at a typical trajectory of the stochastic process as the collapse rate $\gamma$ is progressively increased. The results are shown in Fig. \ref{fig:summary}. One sees that upon increasing the value of $\gamma$, there is a crossover from a continuous diffusion to a jump dynamics (up to some subtleties). Interestingly, the solutions of \eqref{eq:basic} seem to converge, in some non-trivial sense, when $\gamma \rightarrow + \infty$. It is this limit we shall be mostly interested in understanding precisely here.

\section{Setup}

\subsection{The stochastic Schr\"odinger equation and its origin}

We consider a spontaneous collapse model for a quantum state $\ket{\psi}\in\mathscr{H}=\mathds{C}^D$ with $D< +\infty$. The dynamics is postulated to be given by the \emph{stochastic Schr\"odinger equation} (SSE):
\begin{align}\label{eq:sse}
\upd \ket{\psi_t}= \left\{-i H \upd t+  
\sqrt{\gamma}\,\left( \mathcal{O}-\langle \mathcal{O}\rangle_t\right) \upd W_t - \frac{\gamma}{2}\left[\mathcal{O}^\dagger\mathcal{O}- 2 \, \langle \mathcal{O}^\dagger\rangle_t \mathcal{O} + \, \langle \mathcal{O}^\dagger\rangle_t\langle \mathcal{O}\rangle_t\right]\upd t\right\}\ket{\psi_t}  \,, 
\end{align}
where $\mathcal{O}$ is a generic operator\footnote{A non-Hermitian $\mathcal{O}$ can be used to obtain a so called ``dissipative'' collapse model, but we will focus here mostly on the Hermitian case which already yields rich dynamics.}, $\langle \mathcal{O} \rangle_t = \bra{\psi_t} \mathcal{O} \ket{\psi_t}$, $W_t$ is a Wiener process (Brownian motion), $\gamma$ is the collapse strength (or rate), and $H$ is the system Hamiltonian independent of the collapse process. This stochastic differential equation with multiplicative noise is to be understood in the It\^o convention~\cite{oksendal2003}. As an illustration, taking $D$ large and $\mathcal{O}$ to be a discretized version of the position operator $X$, \eqref{eq:sse} would yield an approximation of the ``\emph{Quantum Mechanics with Universal Position Localization}'' (QMUPL) model \cite{diosi1989,bassi2013review} in one space dimension. More complicated setups can easily be considered, where many operators (possibly non-commuting) are being continuously collapsed simultaneously, but we will stick to this simple dynamics \eqref{eq:sse} in what follows.

Where is such a stochastic differential equation coming from? There are at least 3 ways to motivate it:
\begin{enumerate}
    \item \emph{From collapse models in continuous space} -- Starting from a collapse model in continuous space like the continuous spontaneous localization model (CSL), one may derive an effective collapse equation on a smaller Hilbert space. This happens if one considers degrees of freedom that are intrinsically discrete (for example spin in a Bell or EPR experiment), or if only a few states can be reached by the dynamics (for example if the potential $V(\hat{X})$ appearing in the Schr\"odinger equation has a few deep minima).
    \item \emph{From consistency requirements} -- One may ask what is the most general collapse equation that (i) yields a linear evolution for the density matrix averaged over the noise $\bar{\rho}=\mathds{E}[\ket{\psi}\bra{\psi}]$ (ii) is Markovian (iii) preserves state purity. It turns out that all equations with these properties essentially have the same form as \eqref{eq:sse} up to some additional phase factors (see \cite{bassi2013,diosi2014} and references therein).
    \item \emph{From continuous measurement theory} -- Since their inception in the eighties, collapse models have been developed alongside the theory of continuous measurement \cite{jacobs2006,wiseman2009}. The latter aims to describes the continuous monitoring of quantum systems within orthodox quantum theory. In this context, one simply pushes the use of the collapse postulate sufficiently far away from the system studied to avoid problems or ambiguities for all practical purposes. A continuous measurement or monitoring is then obtained in a proper limit where infinitely weak measurements are carried infinitely frequently. It turns out that the equations one obtains in this context are exactly the same as those of continuous collapse models\footnote{This equivalence holds only for Markovian collapse models. For colored noise (or non-Markovian collapse models), there is no longer a simple continuous measurement interpretation.}. The interpretation is of course different, but the formalism is identical. 
\end{enumerate}
This latter motivation from continuous measurement theory is crucial for us, as it provides most of the intuition for the results. Furthermore, considering a finite dimensional Hilbert space is more common and natural on the continuous measurement side, where the systems monitored are typically effective qubits or few level systems, not fundamental constituents of nature. Finally, whilst the stochastic trajectories of \eqref{eq:sse} are not observable in the collapse context, they can be reconstructed in the continuous measurement context where the noise is knowable a posteriori, as it is a function of the (random) measurement results. 

\begin{figure}
    \centering
    \includegraphics[width=0.85\textwidth]{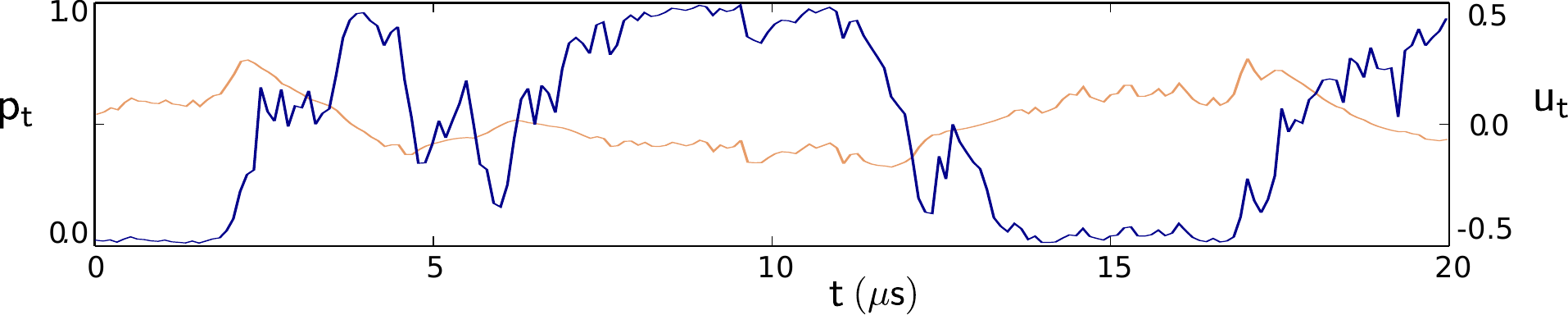}
    \caption[short caption]{\textbf{Quantum trajectory of a continuously monitored transmon qubit} (from \emph{real} experimental data). In blue, $p_t$ is the population in the $z$ basis, and in light orange $u_t=u^*_t$ is the non-diagonal coefficient of $\rho$ in the $z$ basis (see \eqref{eq:param}). The stochastic master equation (see \ref{sec:sme}) describing the evolution is slightly more complicated than the idealized ones we consider in subsequent examples and reads \cite{ficheux2018}:
    \begin{equation*}
    \upd \rho_t = -i[\Omega \sigma_y,\rho_t]\upd t + \sum_j \mathcal{D}[L_j](\rho_t) \upd t +  \sqrt{\eta_d} \, \frac{\Gamma_d}{2} \mathcal{H}[\sigma_z](\rho_t)\upd W_t, 
    \end{equation*}
    with $j= \{u,v,w,\varphi \} $, $L_u= \sqrt{\Gamma_1/2} \sigma_-$, $ L_v= i \sqrt{\Gamma_1/2} \, \sigma_-$, $L_w=\sqrt{\Gamma_d/2}\, \sigma_z$, $L_\varphi=\sqrt{\Gamma_\varphi/2}\,  \sigma_z$, $\Gamma_d = (0.9 \mu s)^{-1}, \Gamma_\varphi = (17.9 \mu s)^{-1}, \Omega = 2 \pi / (5.2 \mu s), \Gamma_1 = (765.3 \mu s)^{-1}$ and $\eta_d = 34 \%$. Experimental data courtesy of Benjamin Huard and Quentin Ficheux of \'Ecole Normale Sup\'erieure. For more detail, see Ficheux's thesis \cite{ficheux2018PhD}. } 
    \label{fig:realtrajectory}
\end{figure}

\subsection{The stochastic master equation}\label{sec:sme}

In practice it is more convenient to work with the equation for $\rho_t= \ket{\psi_t}\bra{\psi_t}$, which makes the general structure more manifest. Using the It\^o formula\footnote{In this context, ``using the It\^o formula'' simply means writing \[ \upd \rho_t = \upd \ket{\psi_t}\; \bra{\psi_t} + \ket{\psi_t}\; \upd \bra{\psi_t} + \upd \ket{\psi_t}\;\upd \bra{\psi_t},\] using the formal rule $\upd W_t \upd W_t = \upd t$ and keeping terms of order one in $\upd t$ and $\upd W_t$.}, we get the \emph{stochastic master equation} (SME):
\begin{equation}\label{eq:sme}
    \upd \rho_t = \mathscr{L}(\rho_t)\,\upd t + \gamma \mathcal{D}[\mathcal{O}] (\rho_t)\, \upd t+ \sqrt{\gamma}\,\mathcal{H}[\mathcal{O}](\rho_t) \,\upd W_t
\end{equation}
where $\mathscr{L}(\rho)=-i[H,\rho]$ and we have used the continuous measurement theory notations:
\begin{align}
    \mathcal{D}[\mathcal{O}](\rho) &=\mathcal{O}\rho\mathcal{O}^\dagger - \frac{1}{2} \{ \mathcal{O}^\dagger \mathcal{O},\rho\}\\
    \mathcal{H}[ \mathcal{O}](\rho)&=\mathcal{O}\rho + \rho \mathcal{O} - \tr\left[(\mathcal{O} + \mathcal{O}^\dagger)\rho\right] \rho.
\end{align} 
This SME \eqref{eq:sme} is equivalent with the SSE \eqref{eq:sse} if the initial state $\rho_t$ is pure ($=$ rank $1$), but it is more flexible, allowing $\mathscr{L}$ that are not Hamiltonian flows and do not preserve purity. Its first term $\mathcal{D}[\mathcal{O}]$ is linear and is simply a Lindblad operator: it encodes the decoherence associated with the collapse process and remains upon averaging over the noise. The second term $\mathcal{H}$ is a non-linear map on $\rho$. It is responsible for the collapse\footnote{In the context of continuous measurement theory, it encodes the progressive acquisition of information and is sometimes called the ``stochastic innovation'' term. } and it would disappear upon noise averaging. Indeed, $\bar{\rho}_t=\mathds{E}[\rho_t]$ is straightforward to compute because It\^o integrals against the Wiener process have zero average. The average density matrix $\bar{\rho}$ verifies the master equation (ME):
\begin{equation}
\frac{\upd}{\upd t} \bar{\rho}_t = \mathscr{L}(\bar{\rho}_t) + \gamma \mathcal{D}[\mathcal{O}] (\bar{\rho}_t).
\end{equation}
The fact that it is linear is fundamental and insures the consistency of the collapse model and prevents problems with signalling or the probabilistic interpretation of the quantum state \cite{gisin1989,bassi2015}. This linearity is very natural in the measurement context, where averaging over the randomness of the measurement results is equivalent to tracing over an environment and thus preserves linearity.

\section{Pure collapse}
\begin{figure}
    \centering
    \includegraphics[width=0.8\textwidth]{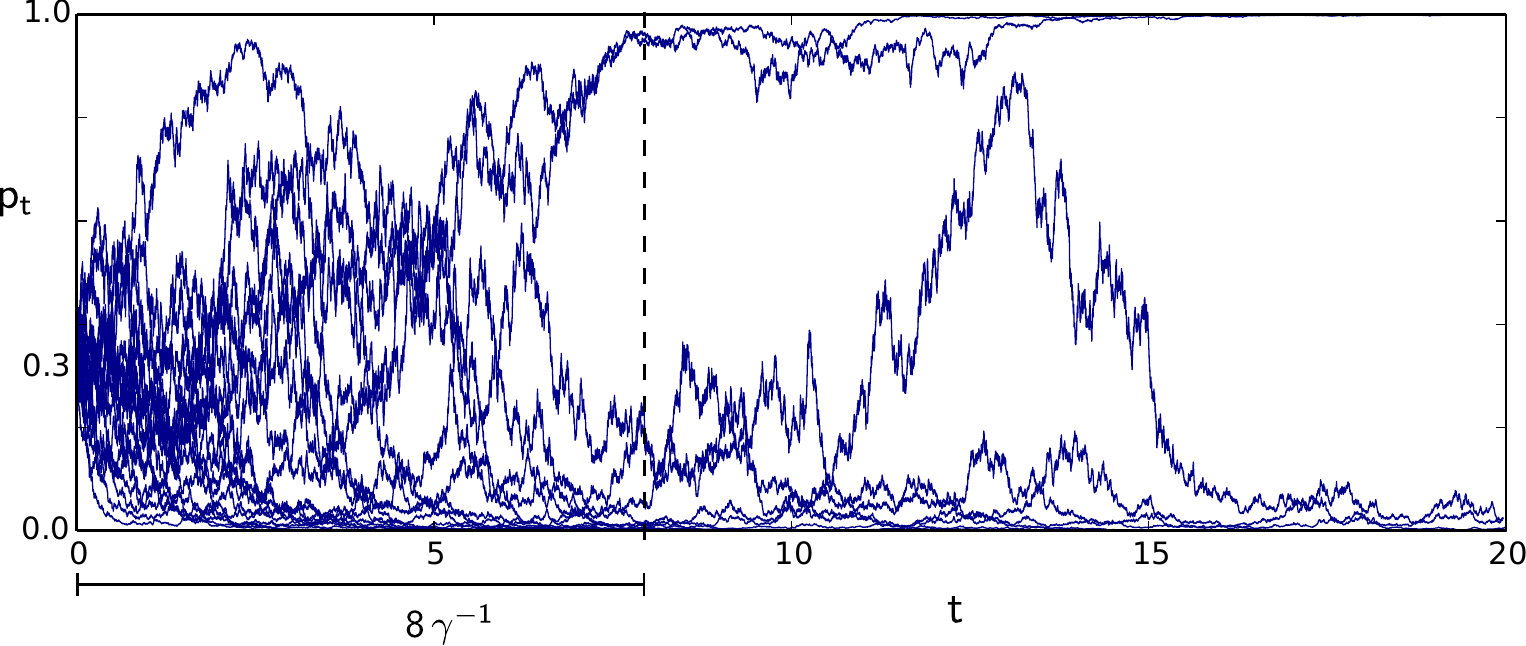}
    \caption{\textbf{A few realizations of the stochastic process} $ [\upd p_t = \sqrt{\gamma} \,  p_t (1-p_t) \,\upd W_t]$ --  for $\gamma =1$ and $p_0=0.3$. Trajectories converge to $p=1$ or $p=0$ exponentially fast on average, with characteristic timescale $\tau \propto \gamma^{-1}$.}
    \label{fig:pure}
\end{figure}

Let us first consider \eqref{eq:sme} in the limit when there is no additional dynamics (\ie $\mathscr{L}=0$) and for a qubit (\ie $\mathscr{H}=\mathds{C}^2$):
\begin{equation}\label{eq:smepuremeasurement}
    \upd \rho_t = \gamma \mathcal{D}[\mathcal{O}] (\rho_t)\, \upd t+ \sqrt{\gamma}\,\mathcal{H}[\mathcal{O}](\rho_t) \,\upd W_t.
\end{equation}
To make things specific, we take $\mathcal{O} = \sigma_z/2$ where $\sigma_x,\sigma_y,\sigma_z$ are the 3 Pauli matrices. We introduce a parameterization of the qubit density matrix:
\begin{equation}\label{eq:param}
    \rho_t = \left(\begin{array}{cc}
         p_t& u_t  \\
         u^*_t& 1-p_t 
    \end{array}\right),
\end{equation}
where $p_t \in [0,1]$ is the probability to be in (or population in) the state $\ket{+}_z$, \ie $p_t=\bra{+}\rho_t\ket{+}_z$, and $u_t$ is a complex phase. We can expand \eqref{eq:smepuremeasurement} to obtain an equation for $p$ and $u$. The one for the phase is
\begin{equation}
    \upd u_t = -\frac{\gamma}{8} u_t \, \upd t +\frac{\sqrt{\gamma}}{2} (2 p_t- 1) u_t \, \upd W_t.
\end{equation}
The stochastic trajectory of the phase depends on the population, but its average obeys an autonomous equation: writing $\bar{u}_t = \mathds{E}[u_t]$ we have $\frac{\upd}{\upd t} \bar{u}_t = -(\gamma/8) \bar{u}_t$, and $\bar{u}_t = \bar{u}_0 \e^{-\gamma t /8}$. Hence, on average, collapse dynamics induces exponential decoherence in the eigenbasis of the collapse operator. This is expected. The equation for $p_t$ is more interesting:
\begin{equation}\label{eq:measure}
    \upd p_t = \sqrt{\gamma} \,  p_t (1-p_t) \,\upd W_t.
\end{equation}
It contains no deterministic part and is a pure ``martingale'' (\ie unbiased on average). A few realizations of this stochastic process are plotted in Fig. \ref{fig:pure}. The diffusion is inhomogeneous and makes $p_t$  converge exponentially fast\footnote{The probability $p_t$ never touches $0$ or $1$ exactly. This behavior is to be contrasted from that of the Wright-Fisher equation encountered in population dynamics. The latter is similar but for a crucial square root:
\begin{equation}
    \upd x_t = \sqrt{x_t(1-x_t)} \, \upd W_t,
\end{equation}
and with this modification the boundaries $x=0$ or $x=1$ are reached almost surely in finite time.} (on average) to $0$ or $1$. This is easily seen by considering $\Delta_t = \sqrt{p_t(1-p_t)}$ which measures the distance from the final state. Using the It\^o formula, one obtains:
\begin{equation}
    \upd \Delta_t = -\frac{\gamma}{8} \Delta_t \, \upd t +\frac{\sqrt{\gamma \Delta_t}}{2}(1-2p_t)\,\upd W_t.
\end{equation}
Hence, $\frac{\upd}{\upd t} \bar{\Delta}_t = -(\gamma/8) \bar{\Delta}_t$, and $\bar{\Delta}_t = \bar{\Delta}_0\,  \e^{-\gamma t /8}$. 

Now we may wonder if the collapse towards $p=1$ or $p=0$ is done according to the Born rule. This is indeed the case:
 \begin{equation}\label{eq:bornrule}
 \mathds{P}\left[\ket{\psi_t}\underset{t\rightarrow+\infty}{\longrightarrow} \ket{+}_z\right]= |\langle \psi_0 \ket{+}_z|^2 \;\;\text{or, equivalently}\;\;
     \mathds{P}\left[p_t\underset{t\rightarrow+\infty}{\longrightarrow} 1 \right] = p_0.
 \end{equation}
This is seen easily by exploiting the fact that $p_t$ is a martingale, a property which explains most of the features of the collapse process (see \eg \cite{adler2001,bauer2011}). Using \eqref{eq:measure}, we have simply that $\frac{\upd \bar{p}_t}{\upd t}=0$, hence $\bar{p}_t \equiv p_0$, and $\lim_{t\rightarrow +\infty} \bar{p}_t = p_0$. The later limit is:
\begin{equation}
   \lim_{t\rightarrow +\infty} \bar{p}_t = \mathds{P}\left[p_t\underset{t\rightarrow+\infty}{\longrightarrow} 1 \right] \times 1 + \mathds{P}\left[p_t\underset{t\rightarrow+\infty}{\longrightarrow} 0 \right] \times 0 = \mathds{P}\left[p_t\underset{t\rightarrow+\infty}{\longrightarrow} 1 \right],
\end{equation}
where we have used the fact that $p_t$ converges almost surely to $0$ or $1$. 

Hence the effect of our stochastic process is (i) to shrink the non-diagonal coefficients in the basis of $\mathcal{O}$, i.e. \emph{decohere} (ii) make the diagonal coefficients all converge towards one, i.e. on has a \emph{collapse} as expected. Both happen exponentially fast, with a rate controlled by $\gamma$, and with the expected probability. This generalizes trivially from $\mathscr{H}=\mathds{C}^2$ to $\mathds{C}^D$ for $D<\infty$.

\section{Jumps} \label{sec:jumps}
Now that we understand the dynamics induced by a continuous collapse process, we can add intrinsic dynamics of the system. In the limit where this dynamics is slow compared to the continuous collapse process, we will see the emergence of quantum jumps. This may be understood as a form of semiclassical limit: how does a quantum system behave when collapse is so fast that the state is almost always well localized?

\subsection{Qubit with dissipative dynamics}
The simplest setup we can consider is that of a qubit coupled to a thermal bath which induces a relaxation in the energy basis. Namely, we consider the evolution:
\begin{equation}\label{eq:smethermal}
    \upd \rho_t = \mathscr{L}_\text{thermal}(\rho_t)\,\upd t+ \gamma \mathcal{D}[\mathcal{O}] (\rho_t)\, \upd t+ \sqrt{\gamma}\,\mathcal{H}[\mathcal{O}](\rho_t) \,\upd W_t.
\end{equation}
with the Lindblad operator:
\begin{equation} 
    \mathscr{L}_\text{thermal}(\rho) = \lambda_\uparrow \mathcal{D}[\sigma_+](\rho) + \lambda_\downarrow \mathcal{D}[\sigma_-](\rho),
\end{equation}
where $\sigma_+ = \ket{+}\bra{-}_z = \sigma_-^\dagger$ and $\lambda_{\uparrow/\downarrow}$ represent the excitation and de-excitation induced by the bath.
The complete evolution preserves diagonal density matrices: as in the pure collapse case, the non-diagonal coefficients shrink exponentially without any feedback on the diagonal ones. As a result, we consider only the evolution of $p_t$.
Expanding \eqref{eq:smethermal} yields:
\begin{align}\label{eq:probathermal}
    \upd p_t =\lambda \, (p_\text{eq} - p_t)\,\upd t + \sqrt{\gamma}\,  p_t (1-p_t) \,\upd W_t,
\end{align}
with $\lambda = \lambda_\downarrow+\lambda_\uparrow$ and $p_\text{eq}=\lambda_\downarrow/\lambda$. This is the equation we advertised in the introduction with its non-trivial competition between collapse driving $p_t$ to $0$ or $1$ and relaxation driving it to $p_\text{eq}$.

For large $\gamma$, the stochastic process $p_t$ seems to converge (in a weak sense) to a Markovian jump process (see \ref{fig:summary}). Intuitively, the dominant noise term $\sqrt{\gamma} \, p_t(1-p_t) \upd W_t$ forces $p_t$ to be almost always $0$ or $1$ and the subleading deterministic term induces Markovian transitions between these boundary values.

For this simple example, one can easily characterize the emerging jump process quantitatively provided one accepts that the large $\gamma$ limit is indeed a Markov process. Such a Markov process would be characterized by two jump rates $M_{(+)\leftarrow (-)}$ and $M_{(-) \leftarrow (+)}$. In the large $\gamma$ limit, $p_t$ becomes a Markov chain between $0$ and $1$ and we its average value $\bar{p}_t$ will thus obey:
\begin{equation}
    \frac{\upd}{\upd t} \bar{p}_t = -M_{(-)\leftarrow (+)} \bar{p}_t + M_{(+) \leftarrow (-)} (1-\bar{p}_t)
\end{equation}
But using \eqref{eq:probathermal} we have that for all $\gamma$ (and not just $\gamma$ infinite):
\begin{equation}
    \frac{\upd}{\upd t} \bar{p}_t = -\lambda_\uparrow p_t +\lambda_\downarrow (1-\bar{p}_t).
\end{equation}
Hence we simply read that $M_{(-)\leftarrow (+)}=\lambda_\uparrow$ and $M_{(+) \leftarrow (-)} = \lambda_\downarrow$. In this very simple example, the jump rates can be read straightforwardly from the averaged master equation and it is only the very emergence of the jump process that is less trivial and requires the stochastic description. 

\subsection{Qubit with coherent dynamics}\label{sec:qubitcoherent}

We now consider a second example where continuous collapse competes with a non-commuting unitary evolution. Namely, we choose a Hamitlonian $H=(\omega/2) \sigma_y$ while still collapsing with the operator $\mathcal{O}=\sigma_z/2$. The SME reads
\begin{equation}\label{eq:smeunitary}
    \upd \rho_t = - i \,\frac{\omega}{2} \, [\sigma_y,\rho_t]\,\upd t+ \gamma \mathcal{D}[\mathcal{O}] (\rho_t)\, \upd t+ \sqrt{\gamma}\,\mathcal{H}[\mathcal{O}](\rho_t) \,\upd W_t.
\end{equation}
As before, it can be expanded into a pair of stochastic differential equations for $p_t$ and $u_t$ which parameterize the density matrix. However, this time the two equations are coupled, and the non-diagonal coefficient $u_t$ does not shrink to zero. In fact, real and pure density matrices are preserved by the evolution \eqref{eq:smeunitary} so that there is still only one dynamical parameter (which is an angle in the Bloch sphere). But the discussion remains easier with $p_t$ and $u_t$. 

\begin{figure}
    \centering
    \includegraphics[width=0.85\textwidth]{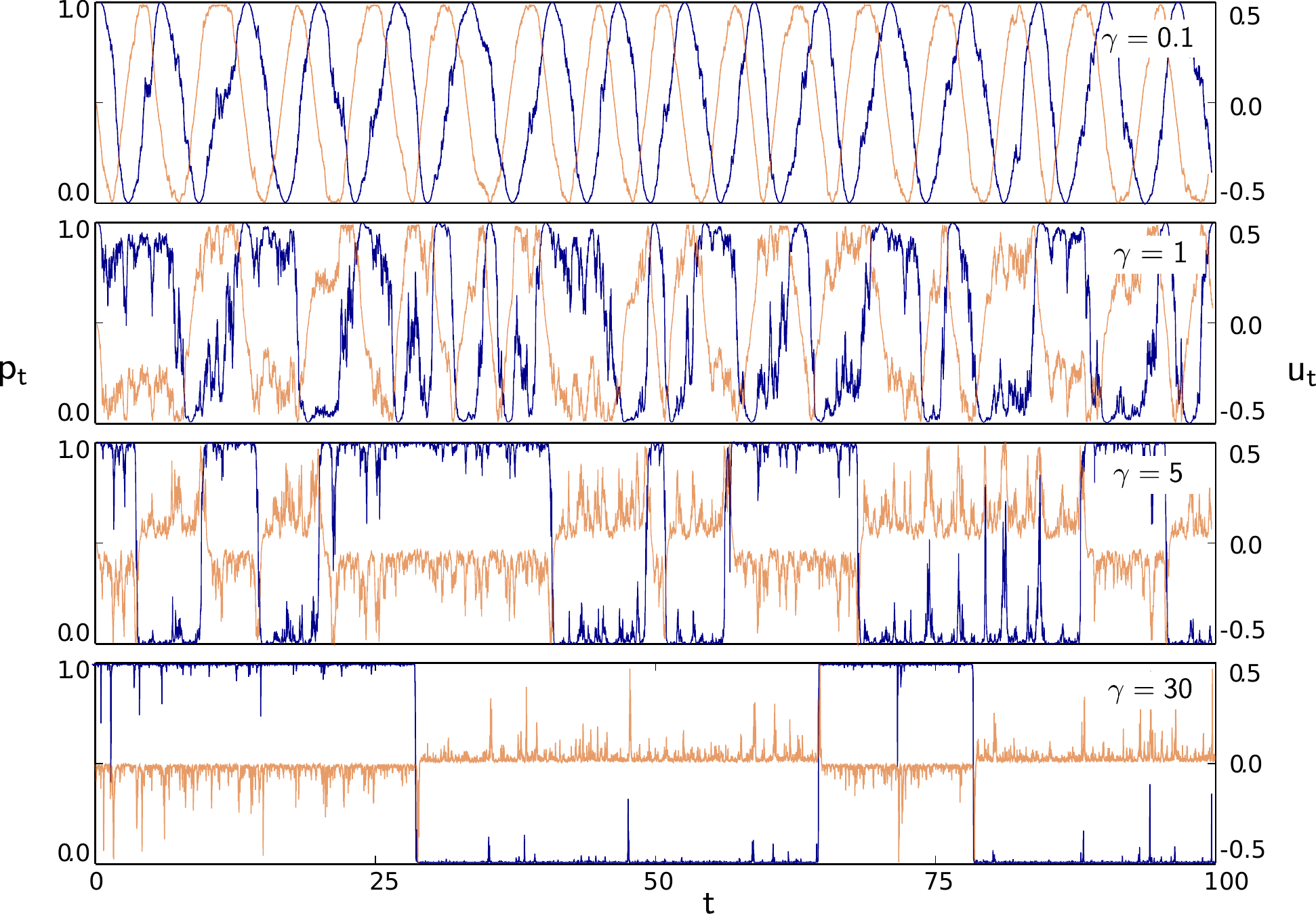}
    \caption{\textbf{Trajectories $p_t$} (dark blue)  \textbf{and} $u_t$ (light orange) from   -- Typical trajectories are shown for $\omega = 1$, $p_\text{eq}=1.0$, and $\gamma = \{0.1, 1, 5, 30\}$. They are obtained through a naive Euler discretization of \eqref{eq:basic} with $\upd t = 10^{-4}$. $u_t$ is real. For $\gamma \gg 100$ we would no longer see any jump in the figure and $p_t$ would appear stuck in either $0$ or $1$.}
    \label{fig:unitarysummary}
\end{figure}

Intuitively, what do we expect will happen? In the $z$ basis, the unitary evolution with $\sigma_x$ creates Rabi oscillations, hence $p_t \sim \cos(\omega t)$ in the absence of collapse. On the other hand, when continuous collapse dominates, we expect $p_t$ to spend most of its time near $0$ or $1$ as before. There is indeed some non-trivial competition between the two. Let us just see how the trajectories look in Fig. \ref{fig:unitarysummary}. As before, we observe an emergent jump behavior in the large $\gamma$ limit, starting from a completely different evolution for $\gamma$ small (Rabi oscillations versus thermal relaxation in the previous case). However, there an important difference: as $\gamma$ increases, the jumps get sharper and more discontinuous, but at the same time their frequency decreases (in $1/\gamma$). This is a signature of the Zeno effect: a coherent transition is slowed down by collapse (or measurement).

In this example, one can also prove rigorously the emergence of jumps and compute their rate explicitly. However, it is easier to go directly to the general case, which makes the perturbative reasoning more transparent.

\subsection{General case}

\begin{figure}
\begin{center}
    \includegraphics[width=0.5\textwidth]{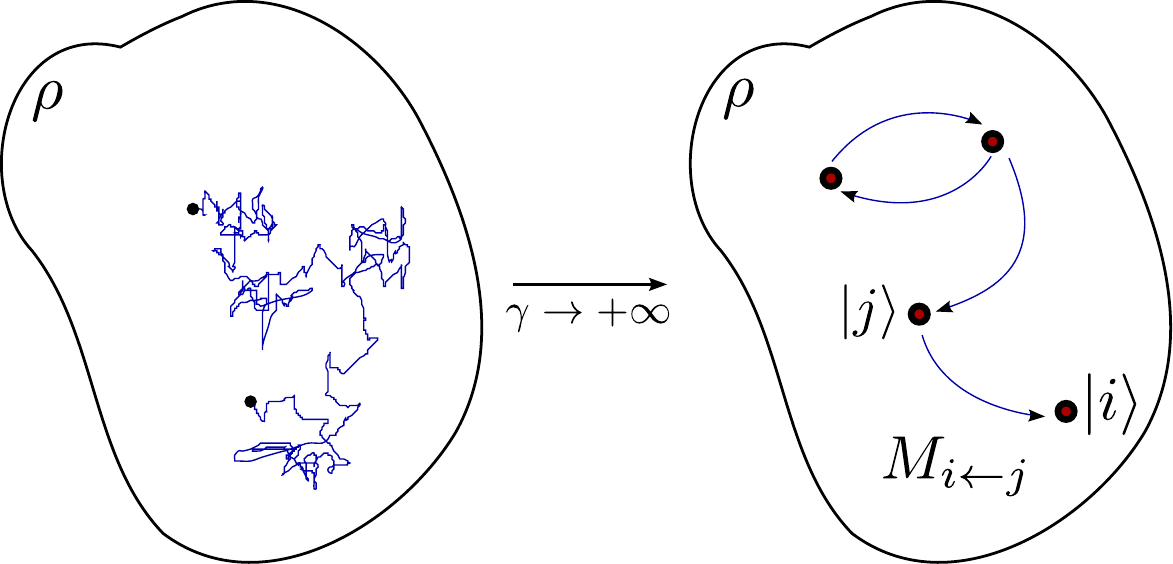}
    \caption{\textbf{Illustration of the jump theorem} -- For finite $\gamma$, the state $\rho$ diffuses in Hilbert space. When the collapse rate is sent to infinity, the state spends most of its time near the eigenvectors of the collapse operator. It becomes a Markov chain, randomly jumping from pointer to pointer with a rate $M_{i\leftarrow j}$ that can be computed explicitly.}
    \label{fig:jumpthm}
\end{center}
\end{figure}

We saw in the two examples before that the emergence of jumps seemed ubiquitous in the strong measurement limit, but that their rate depends on their origins: jumps mediated by a bath have a fixed rate independent of the measurement strength whereas ``unitary'' jumps are Zeno suppressed. For the latter, the jump rates vanish for large $\gamma$, and would thus be zero if the limit were taken too brutally. It means that to obtain a non-trivial limit, we need to adequately rescale the system dynamics (given by the Liouvillian $\mathscr{L}$) while the measurement strength $\gamma$ is sent to $+\infty$. Up to this subtlety, we will show, or rather suggest, that the jump limit is ubiquitous, and there is generically a transition from continuous diffusive dynamics to discrete jump dynamics in the fast collapse limit (see Fig. \ref{fig:jumpthm}).

We recall the setup, following the derivation in \cite{bauer2015}. We continuously collapse a certain self-adjoint operator $\mathcal{O} = \sum_{k} \nu_k \ket{k}\bra{k}$ at a rate $\gamma$, where the $\nu_k$ are real and, we assume, all different. We have a system evolution in the absence of collapse given by $\mathscr{L}_\gamma$ which depends on $\gamma$ because we allow ourselves to rescale the part of the dynamics yielding jumps that would otherwise be Zeno suppressed. The evolution of the density matrix reads
\begin{equation}\label{eq:generaljumpevol}
    \upd \rho_t = \mathscr{L}_\gamma(\rho_t)\,\upd t+ \gamma \mathcal{D}[\mathcal{O}] (\rho_t)\, \upd t+ \sqrt{\gamma}\,\mathcal{H}[\mathcal{O}](\rho_t) \,\upd W_t.
\end{equation}
We now need to parameterize the Liouvillian more explicitly. To this end, we write $[\mathscr{L}(\rho)]^{ij} = L^{ij}_{kl}\rho^{kl}$ with summation on repeated indices and postulate the scaling:
\begin{equation}\label{eq:scaling}
    \begin{split}
        L^{ii}_{ll} &= A^{i}_{l} + o(1)\\
        L^{ii}_{kl} &= \sqrt{\gamma} \, B^{i}_{kl} + o(\sqrt{\gamma}) \;\; \text{for} \;\; k\neq l \\
        L^{ij}_{ll} &= \sqrt{\gamma} \, C^{ij}_l + o(\sqrt{\gamma}) \;\; \text{for} \;\; i\neq j \\
        L^{ij}_{kl} &= \gamma \, D^{ij}_{kl} + o(\gamma)\;\; \text{for} \;\; i\neq j \;\; \text{and} \;\; k\neq l \;\; \text{and} \;\; D^{ij}_{kl} = -d_{kl} \delta^{i}_{k} \delta^{j}_l.\\
    \end{split}
\end{equation}
The justification for this scaling comes naturally when calculating the jump rates. Since we shall not carry the proof here, let us just say that the $A$ term corresponds to incoherent contributions like those of the first example, and thus has to be taken fixed. The $B$ and $C$ terms essentially correspond to a Hamiltonian contribution like that of the second example, and need to be enhanced as $\gamma$ is increased to get a non-zero jump rate. Finally, the diagonal part of the $D$ term has an effect similar to that of the collapse on the average density matrix, and thus needs to be scaling like $\gamma$ to remain relevant in the limit.

The jump theorem then gives \cite{bauer2015} (see also \cite{ballesteros2019}) that in the large $\gamma$ limit, $\rho$ becomes a Markov chain between the projectors $\ket{k}\bra{k}$ with jump rates (or Markov matrix):
\begin{equation}\label{eq:thm1}
    M_{i \leftarrow j}=A^i_j +2\, \Re \mathrm{e} \sum_{k<l} \frac{B^i_{kl} C^{kl}_j}{\Delta_{kl}}
\end{equation}
with $\Delta_{kl}=\frac{1}{2}|\nu_k-\nu_l|^2  + d_{kl}$. To give some intuition about the result we consider a slightly less general situation where $\mathscr{L}(\rho) = A (\rho) - i[H,\rho]$ with $A$ acting diagonally as in \eqref{eq:scaling}. Then the jump rates simplify to:
\begin{equation}
M_{i\leftarrow j} =\hskip-0.9cm \overset{\text{ ``incoherent'' contribution}}{\overbrace{{L}^{ii}_{jj}}} \hskip-0.4cm + \underset{\text{``coherent'' contribution}}{\underbrace{\frac{4}{\gamma}\left|\frac{H_{ij}}{\nu_i-\nu_j}\right|^2}}.
\end{equation}
So again, if $H$ is not rescaled $\propto \sqrt{\gamma}$, the coherent contribution is suppressed in the limit. Note the interesting form of this coherent term: it depends not only on the collapse basis but also on the eigenvalues of the collapse operator. This is a very distinct behavior from the one obtained from a projective measurement or instantaneous collapse to a pointer, where nothing physical can depend on the eigenvalues.

There are two strategies to derive \eqref{eq:thm1}, a quick and dirty method using the master equation, and a more rigorous one using the evolution for the probability distribution of $\rho$:
\begin{enumerate}
    \item One can accept, from the pure collapse discussion, that the collapse will make the state stay near the eigenvectors of the collapse operator most of the time, and further assume the transitions from pointer to pointer will be Markovian. Then, to compute their rate, one only needs to study the master equation for $\bar{\rho}$ (averaged over the collapse noise):
    \begin{equation}\label{eq:sdecompletemoy}
    \partial_t \bar{\rho}_t = \left(A + \sqrt{\gamma} \left(B + C\right) + \gamma \left(D+\mathcal{D}[\mathcal{O}]\right) \right)(\bar{\rho}_t).
\end{equation}
    On then carries perturbation theory to second order in $\gamma$ to find a closed master equation for the diagonal part of $\rho$: $\partial_t\text{diag} (\bar{\rho}_t) = M\, \text{diag}(\bar{\rho}_t)$, where $\text{diag} (\bar{\rho}_t)$ is written as a column vector. The matrix $M$ is then identified as the Markov matrix of \eqref{eq:thm1}. This is the strategy followed in \cite{tilloy2016phd}: it is simple as one only needs the master equation (and not the stochastic master equation) but requires one to assume that the limit is indeed a Markov process between pointers\footnote{To see that this is not obvious, note that there exist other unravelings of the master equation \eqref{eq:sdecompletemoy}, \ie different stochastic master equations giving the same average master equation, that do not give jumps between pointers in the limit. Hence the jump limit really is a feature of the stochastic description.}.
    \item A more rigorous method consists in going one step more abstract. The idea is to study not only the average of the state $\bar{\rho}$, but rather its full probability distribution $\mathds{P}_t[\rho\, |\, \rho_0]$. From the stochastic master equation \eqref{eq:generaljumpevol}, one can find the second order Fokker-Planck operator $\mathfrak{D}$ such that 
    \begin{equation}
        \partial_t \mathds{P}_t[\rho\, |\, \rho_0]=  \mathfrak{D} \mathds{P}_t[\rho\,|\,\rho_0].
    \end{equation}
    With the scaling we have chosen, this differential operators admits the expansion $\mathfrak{D} = \mathfrak{D}_0 + \gamma \mathfrak{D}_1$, hence $\mathds{P}_t = \exp\left(t\mathfrak{D}_0 + t\gamma \mathfrak{D}_1\right) \mathds{P}_0$. One then notes that for large $\gamma$, probability distributions which survive are in the kernel of $\mathfrak{D}_1$ because it is a negative operator. One then shows that these probability distributions are Dirac measures on the eigenvectors of the collapse operator. A perturbative expansion around these stable points in the space of probability distributions gives the Markovian transitions between them \cite{bauer2015}. Hence this method allows to prove that the large $\gamma$ limit is indeed that of a Markovian jump process between pointers at the same time as it allows to compute the jump rates.
\end{enumerate}
In what sense do we have convergence towards the jump process in the large $\gamma$ limit? Actually, we have no more than a convergence in law, that is, expressions of the form $\mathds{E}[f(\rho_{t_1},...,\rho_{t_N})]$ converge to the ones computed with the limiting jump process in the large $\gamma$ limit. Could we hope to prove more? No, because a fine grained structure, that is not captured by the jump process, survives in the limit: the \emph{quantum spikes}.

\section{Spikes}

\begin{figure}
    \centering
    \includegraphics[width=0.99\textwidth]{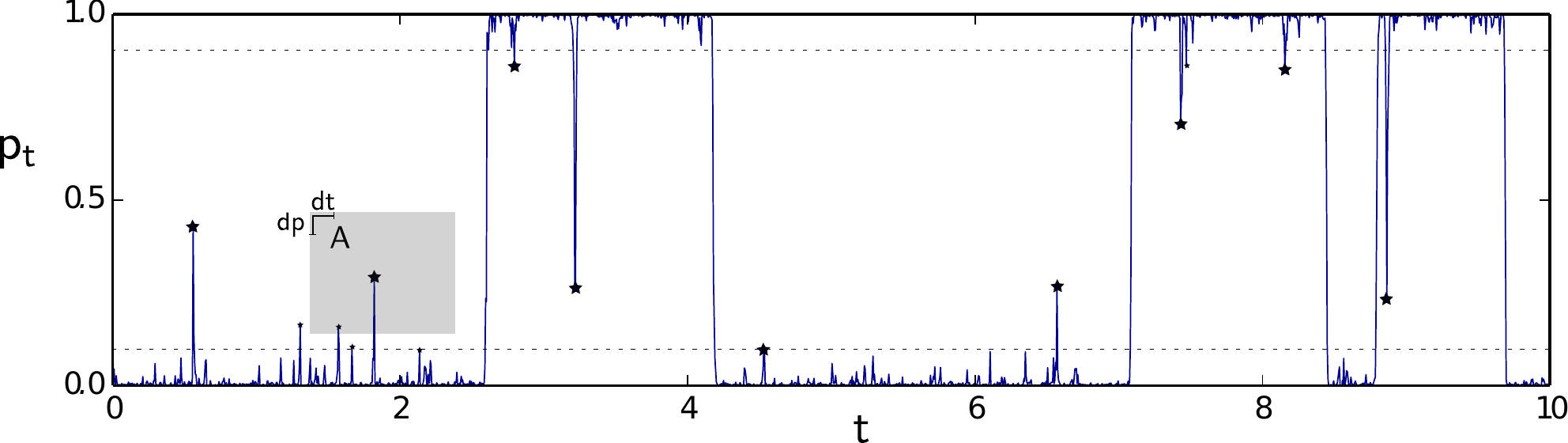}
    \caption{\textbf{Trajectories $p_t$ for $\gamma \gg \lambda$} -- A typical trajectory for $\lambda = 1$, and $\gamma = 400$, is visually very close to what one would get for $\gamma = +\infty$. Spikes above a certain threshold (here $10\%$) are marked with stars. To quantify spikes, one considers a domain $A$ of the plane $(t,p)$ in a region without jumps. The number of spikes ending in $A$ is a Poisson random variable with an intensity given in \eqref{eq:formulaspikes}.}
    \label{fig:spikes}
\end{figure}
\subsection{A first observation}
To understand the phenomenon of spikes, we will restrict our analysis to the simplest instance in which they appear, in the context of the scalar stochastic differential equation \eqref{eq:probathermal}
\begin{equation}
        \upd p_t =\lambda \, (p_\text{eq} - p_t)\,\upd t + \sqrt{\gamma}\,  p_t (1-p_t) \,\upd W_t.
\end{equation}
Already from Fig. \ref{fig:summary}, the careful reader will have noticed that $p_t$ does not quite converge to a jump process. There seem to be sharp excursions decorating the jump process which one would almost dismiss as numerical artifacts. We call these seemingly instantaneous excursions \emph{quantum spikes} and highlight them them in Fig. \ref{fig:spikes}.

Quantum spikes are fast in the sense that they take a time $\propto \gamma^{-1}$ and thus appear discontinuous when $\gamma \rightarrow + \infty$. However, their height remains of order $1$ in the limit. Thus, while they disappear in the sense of Lebesgue measure in the fast collapse limit (and thus in quantities like $\mathds{E}[f(p_{t_1},...,p_{t_N})]$), they remain if one considers instead first passage times or statistics of local extrema. 

\begin{figure}
    \centering
    \includegraphics[width=0.6\textwidth]{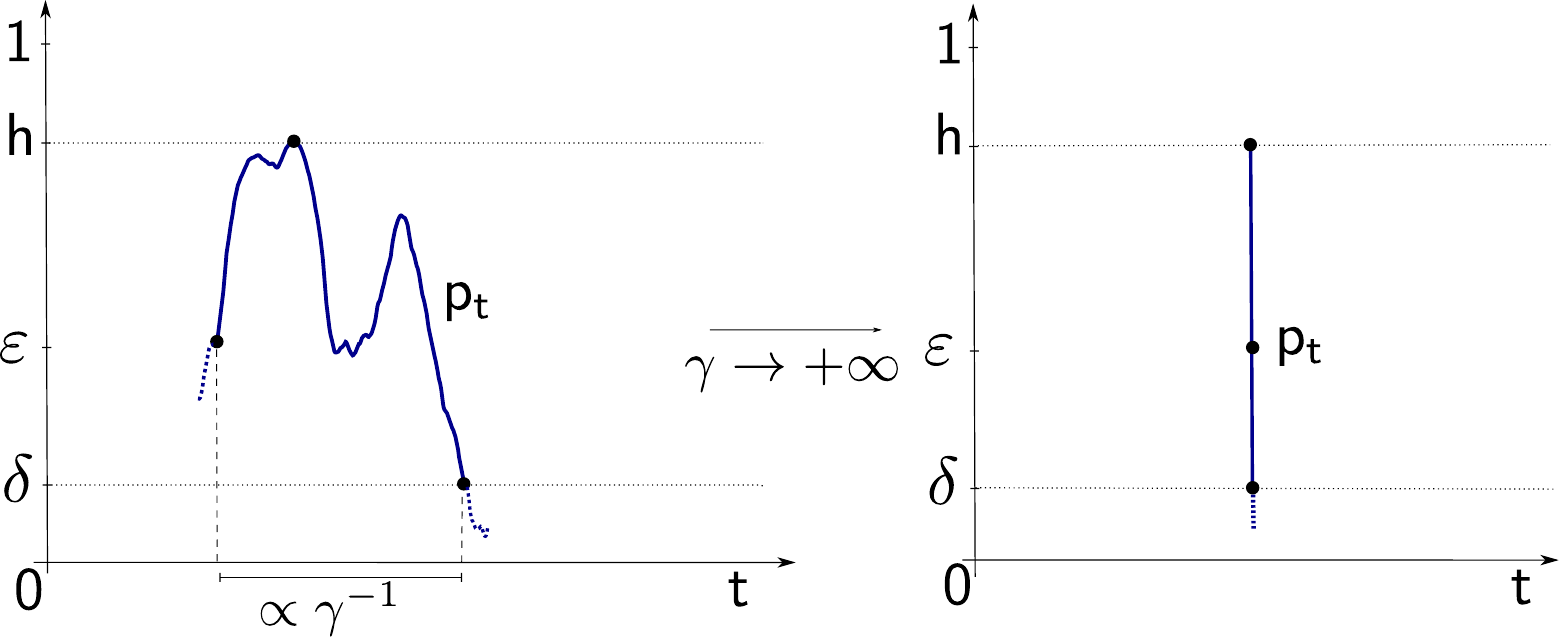}
    \caption{For $\gamma$ finite, one needs two thresholds $0<\delta < \varepsilon \ll 1$ to define a spike (starting from $0$). We consider that an excursion starts when the process reaches $\varepsilon$ from below, and then stops when it hits $\delta$. In the middle, the excursion reached a maximum here written $h > \varepsilon$. When $\gamma\rightarrow + \infty$, the excursion becomes instantaneous, all that remains is a vertical line up to $h$, which we call a spike. Once $\gamma$ has been sent to $+\infty$, the two thresholds $\varepsilon$ and $\delta$ can be sent to zero, and spikes of arbitrarily small size may be considered.}
    \label{fig:spikedef}
\end{figure}

It is rather obvious to see what a spike is from a plot like that of Fig. \ref{fig:spikes}, but it is important (and less trivial) to define spikes more precisely. For simplicity, we consider upward spikes starting from $0$ (downward spikes, starting from $p=1$ are treated in the same way). Let us give ourselves two fixed thresholds $\delta \ll \varepsilon\ll 1$. We call a spike an excursion, or piece of trajectory, starting from $\varepsilon$ and eventually reaching $\delta$. Because excursions away from $0$ become instantaneous in the large $\gamma$ limit, the only thing we see from them is a vertical line from $\varepsilon$ up to the maximum value reached during the excursion and down to $\delta$, hence the name spike (see Fig. \ref{fig:spikedef}). Once we have sent $\gamma$ to $+\infty$ and spikes are effectively instantaneous, we can lower the thresholds $\varepsilon$ and $\delta$ arbitrarily close to $0$ so that the statistics of spikes do not depend on them. 

With this definition, spikes can be given a precise characterization. Namely, the number of spikes ending in a finite domain $A$ of the plane $(p,t)$ (see Fig. \ref{fig:spikes}) is a \emph{Poisson process} of intensity $\mu$:
\begin{equation}
   \mathds{P}\left[n \; \text{spikes ending in}  A\right]=\frac{\e^{-\mu}\mu^n}{n!}\;\; \text{with}\;\; \mu= \int_A \upd \nu(p,t)
\end{equation}
The density $\nu$ is then given by the following (truncated) power laws:
\begin{align}\label{eq:formulaspikes}
    \upd \nu_0(p,t) &= \upd t \, \upd p \, \frac{\lambda \, p_{eq}}{p^2} \;\; \text{for spikes starting from } 0\\
    \upd \nu_1(p,t) &= \upd t \, \upd p \, \frac{\lambda \, (1-p_{eq})}{(1-p)^2} \;\; \text{for spikes starting from } 1.
\end{align}
Importantly, $\gamma$ appears nowhere, the limiting distribution is well defined for $\gamma$ infinite. Further the integrated density diverges for small spikes $\int_{]0,\varepsilon]\times\Delta t} \upd\nu_0(p,t) = +\infty$, and there are thus infinitely many of them.

\subsection{Martingale intuition}
The essence of the reason for the existence of spikes is the following:
\begin{quote}
As the process $p_t$ is a martingale away from the boundaries when $\gamma\rightarrow + \infty$, if there are jumps, there must be aborted jumps (or spikes) as well. 
\end{quote}
Let us make this argument more precise. Away from the boundaries $p=0$ or $p=1$, and when $\gamma$ is large, the collapse term dominates and $\upd p_t \simeq \sqrt{\gamma} \, p_t (1-p_t) \,\upd W_t$. Importantly, this means that $p_t$ is approximately a martingale, in particular:
\begin{equation}\label{eq:martingaleequality}
    \forall T\geq t,\; \mathds{E}\big[\,p_T \, | \,  p_t=\varepsilon\, \big] \simeq \varepsilon.
\end{equation}
Let us imagine the process started near $0$ at $t_0$ and has increased to a small value $p_{t_0}=\varepsilon$. What is the probability that the jump completes before $p$ goes below $\delta$? Let us write $\tau\geq t$ the time when $p$ hits $\delta$ (and the jump is considered aborted) or hits $1-\delta$ (and the jump is considered completed). This random variable is a so called stopping time, which implies that equation \eqref{eq:martingaleequality} holds if $T$ is replaced by $\tau$ \cite{oksendal2003}. Furthermore, by definition we have:
\begin{equation}
    \mathds{P}[\text{jump completes}| p_{t_0}=\varepsilon ] \times (1-\delta) + \mathds{P}[\text{jump aborts} | p_{t_0}=\varepsilon] \times \delta = \mathds{E}[p_\tau|  p_{t_0}]. 
\end{equation}
The latter term is just $p_{t_0}=\varepsilon$ because the process is a martingale. Hence \begin{equation}
    \mathds{P}[\text{jump completes}| p_{t_0}=\varepsilon] = \varepsilon + \text{negligible corrections } \mathcal{O}(\delta),
\end{equation}
\ie \emph{the probability that a jump completes is equal to how far it already went!} In turn, this means there are jumps that do not complete, excursions that reach a certain value $p$ (for example $1/2$), and then go back to their initial value. 

\subsection{Sketch of a proof}

The previous ``martingale'' argument for the existence and even necessity of spikes is almost sufficient to compute their distribution. As before, we look only at spikes starting near $0$, and we neglect subleading terms $\mathcal{O}(\delta)$. However, instead of looking at the jump completion, we can look at the stopping time $\tau_h$ for a given fixed height $h$ such that $\varepsilon <h < 1$. This random variable just gives the time when $p$ reached $h$ or the $\delta$-neighbourhood of $0$ after starting in $\varepsilon$. We have:
\begin{equation}
    \mathds{P}[p_{\tau_h}=h |  p_{t_0}=\varepsilon ] \times h + \mathds{P}[p_{\tau_h}=0 |  p_{t_0}=\varepsilon] \times 0 = \mathds{E}[p_{\tau_h}|  p_{t_0}=\varepsilon] = \varepsilon. 
\end{equation}
Hence,
\begin{equation}
     \mathds{P}[p_{\tau_h}=h |  p_{t_0}=\varepsilon ] = \frac{\varepsilon}{h}.
\end{equation}
This probability is also the probability that the maximum $p$ reaches before going back to the $\delta$-neighbourhood of $0$ is superior to $h$:
\begin{equation}
    \mathds{P}\left[\max_{u<\tau} (p_u) \geq h \, | \, p_{t_0}=\varepsilon \right] = \frac{\varepsilon}{h}.
\end{equation}
Therefore, we have in differential form:
\begin{equation}
    \upd\mathds{P}\left[\max_{t_0<u<\tau} (p_u) = h < 1-\delta \, | \,p_{t_0}=\varepsilon \right] = \varepsilon \, \frac{\upd h}{h^2}.
\end{equation}
This explains the $1/p^2$ in the density of spike maxima \eqref{eq:formulaspikes}. Note in passing that there is an additional term for $h=1$ (again, up to a $\delta$ neighbourhood), because $\mathds{P}\left[\max_{u\leq\tau} (p_u) = 1 \, | \, p_{t_0}=\varepsilon \right] = \varepsilon$. Hence 
\begin{equation}
    \upd\mathds{P}\left[\max_{t_0<u\leq\tau} (p_u) = h \leq 1 \, | \,p_{t_0}=\varepsilon\right] = \varepsilon \left[ \frac{\upd h}{h^2} + \delta(1-h)\,\upd h \right].
\end{equation}
How often do we get to try to jump, \ie how often does the process reach at least $\varepsilon$ in any small time interval $\Delta t$? To answer this question, we can use a simple consistency argument, namely that the probability to reach $\varepsilon$ is related to the jump rate which we know from the previous section \ref{sec:jumps}.

Let us consider a time interval $\Delta t$ such that $\varepsilon \lambda^{-1} \ll\Delta t\ll \gamma^{-1}$. The probability that a jump from $0$ to $1$ occurs during $\Delta t$ is simply $\lambda p_\text{eq} \Delta t$. This jump probability can be decomposed into the probability to reach at least $\varepsilon$ and then complete a jump\footnote{Note that we do not consider the probability that $\varepsilon$ is reached more than once during $\Delta t$ and thus that there could be several jump attempts. This is because the probability to reach $\varepsilon$ during $\Delta t$ is already much smaller than $1$ for our choice of $\Delta t$ and thus probabilities of having more than one attempt per $\Delta t$ are subleading.}, which reads, from the previous discussion $\mathds{P}[\{p_u\}_{t\leq u \leq t+ \Delta t} \; \text{reaches at least} \; \varepsilon] \times \varepsilon$. Hence we have
\begin{equation}
    \mathds{P}[\{p_u\}_{t\leq u \leq t+ \Delta t} \; \text{reaches at least} \; \varepsilon] = \frac{\lambda p_\text{eq}}{\varepsilon}\Delta t.
\end{equation}
As a result, in every small interval $\Delta t$, the probability density that there is an excursion reaching a maximum $h$ is
\begin{equation}
\upd\mathds{P}\left[\max_{t<u\leq t+\Delta t} (p_u) = h < 1 \, | \,p_{t_0}=\varepsilon\right] =  \frac{\Delta t \lambda p_\text{eq}} {h^2} \upd h.
\end{equation}
Because $\Delta t \gg \gamma^{-1}$, the statistics of spikes from two different intervals are independent and we thus get that the maxima of the excursions are given by the Poisson process we advertised in \eqref{eq:formulaspikes}.

This sketch of proof, relying almost only on the martingale property, follows  closely the way spikes were first characterized \cite{tilloy2015}. This line of argument has been made rigorous by Kolb \& Lisenfeld \cite{kolb2019}. Other proof strategies exist, notably one exploiting a time reparameterization of the dynamics transforming $p_t$ into a reflected Brownian motion (see \cite{bauer2016} for a physicist explanation of the argument, and \cite{bauer2018,bernardin2018} for mathematically rigorous derivations). The case of the qubit with a coherent evolution discussed in \ref{sec:qubitcoherent} can also be treated, and the spikes have the same power law statistics (up to a different prefactor). However, although spikes show up in numerical simulations of dynamics in larger Hilbert spaces and seem ubiquitous, no theoretical characterization is known beyond the qubit case.

\subsection{Are spikes real?}
Now that we have precisely characterized them and are sure of their mathematical existence, we should ask ourselves whether spikes are relevant. Are spikes real, out there in the world, or just a modeling artifact? In fact, this question is far subtler than it seems, and the answer depends on what ontological commitments one makes.

\paragraph*{Classical spikes and hidden Markov models --}
A first question we could ask is to know if spikes could appear classically, merely as the result of imperfect knowledge of an underlying (well defined) jump process. This is indeed the case for the equation we have focused on 
\begin{equation}\label{eq:classicalspikes}
    \upd p_t = \lambda\, (p_\text{eq}-p_t)\,\upd t + \sqrt{\gamma} \, p_t(1-p_t) \,\upd W_t
\end{equation} 
which can be obtained as the real time probability of a hidden Markov model \cite{tilloy2015}. More precisely, consider a (classical) Markov process $R_t$ ($R$ for real) that can jump randomly from the value $0$ to the value $1$ in such a way that, on average:
\begin{equation}
    \frac{\upd}{\upd t} \bar{R}_t = \lambda (p_{eq} -\bar{R}_t) \;\; \text{with} \;\; \bar{R}_t = \mathds{E}[R_t].
\end{equation}
One can construct a model of (classical) continuous imperfect observation of this process (say with repeated blurry pictures) and consider the \emph{filtered} probability $p^\text{f}_t =\mathds{P}[R_t = 1| \text{blurry pictures up to}\; t]$. This filtered probability $p_t^\text{f}$ encodes the knowledge we have of the Markov process position $R_t$ at time $t$ using all past blurry pictures. There exists a particular classical imperfect observation scheme such that $p_t^\text{f}$ verifies exactly equation \eqref{eq:classicalspikes} \cite{tilloy2015}. Hence, in this context, spikes, which are still somehow unexpected, can be explained as an artifact of our residual ignorance of the underlying jump process. Even with the optimal filter, we cannot know the system arbitrarily well because it jumps instantly and we \emph{know} that it does! At the Bayesian optimum, provided by the filter, one gets (perhaps surprisingly) a lot of false alerts (the spikes).

More generally, if the density matrix remains diagonal in the collapse basis (which is the case for incoherent transitions), there exists a classical hidden Markov model whose filtered probability verifies exactly the stochastic master equation we put forward. In fact, in the continuous measurement context, SMEs can be understood as the generalization to non-diagonal matrices of the Kushner-Stratonovich filtering equations used in the context of classical estimation. Hence, so long as one sticks with diagonal density matrices, one can always interpret the collapse as a Bayesian updating of a state of knowledge about a well defined classical variable that evolves independently of the collapse/measurement (that is, without mechanical back-action). Naturally, this equivalence breaks down if the density matrix is not diagonal during the evolution (for example if it evolves unitarily) as is the case in the second example we considered in \ref{sec:qubitcoherent}.

\paragraph*{Quantum spikes and hidden variable theories --} In the general case, the convenient classical interpretation can no longer explain the spikes away. One could still construct hidden variables\footnote{Such hidden variable theories are easy to construct, and are essentially the discrete version of Bohmian mechanics introduced by Bell in the context of quantum field theory. Including a continuous collapse/measurement on top of such dynamics is done \eg in \cite{tilloy2016phd}.} doing discrete jumps (without spikes) but then their jump probabilities would depend on the quantum state \cite{tilloy2016phd}, and thus the spikes would appear to be physical. Hence, while there are spikes that can be explained classically, it seems there exists genuinely quantum spikes as well, which cannot be dismissed as easily as an artifact of Bayesian updating.

\paragraph*{A matter of ontology --} In the end, to know if spikes would be real or not in the context of a given continuous collapse model, one needs to say precisely what is real in the first place, \ie what the ontology of the theory is. A popular choice is to take some expectation value over the state\footnote{In the context of physically realistic collapse models, the operator is typically position dependent and proportional to the regularized mass density.}, \ie $\bra{\psi_t} \mathcal{O}\ket{\psi_t}$. For such a choice, spikes are unequivocally real. But there are other possibilities, for example flashes (or their continuous equivalent sometimes called ``signal'' in the continuous measurement context). These latter ontologies are convenient in some cases, as they allow to consistently couple the collapse model describing quantum matter with a classical sector (for example gravity \cite{tilloy2018,tilloy2019}). Further, for what interests us here, these ontologies do not have spikes. 

\paragraph{Trimming spikes by knowing the future --} There is a third option, also inspired from continuous measurement theory but which, to our knowledge, has never been considered in the context of collapse models: forward-backward estimates (or rather their quantum version). In the classical case, we saw that a filtered probability
\begin{equation}
   p^f_t := \mathds{P}\left[R_t =1 | \{\text{observations before } t\}\right], 
\end{equation}
encoding the knowledge one has in real time about a classical Markov process had spikes. Another quantity, quite natural in the classical context, is the forward-backward or a smoothed probability:
 \begin{equation}
     p^{f.b.}_t:=\mathds{P}\left[R_t=1 | \{\text{all past and future observations} \}\right].
 \end{equation} 
 This quantity can only be computed after all the observations have been carried, and not in real time. Intuitively, it is easy to know a posteriori that a spike was just a spike and that no real jump was about to happen, and thus the forward-backward probability $p^{f.b.}_t$ should not have spikes. This intuition is confirmed by numerical simulations: $p^{f.b.}_t$ is smoother (differentiable) and without spikes \cite{tilloy2015}. In the quantum context, there is no unambiguous definition of a density matrix conditioned on the past and the future. Different notions have been put forward, like the \emph{past quantum state} \cite{gammelmark2013} and the \emph{smoothed quantum state} \cite{guevara2015,gammelmark2013}. For the former, spikes disappear, but the resulting state generically loses its density operator properties (which does not matter if one just aims to define an ontology from an expectation value) while in the former the output is still a \emph{bona fide} quantum state $\rho^{f.b.}_t$ but spikes are generically not tamed.

\paragraph*{Summary --} The reality of spikes depends on the choices we make. One of the equations we obtained in the quantum context and that shows spikes can be obtained from a (classical) hidden Markov model. In the latter, spikes live in our minds only: nothing is spiky in Nature, but our best \emph{real time} knowledge is. Spikes vanish once we look back and are only asked to tell a posteriori where the process was. This makes it a bit unsatisfying to have the spikes be real in the quantum context, especially if it implies giving them a reality as well in the cases that could be just as well described by a classical hidden Markov model. This very minor aesthetic criterion could help compare different collapse model ontologies.

\section{Generalization and open problems}
Let us summarize what is known on the mathematical front. For a finite dimensional Hilbert space and a generic continuous collapse process, one can easily prove a convergence towards pointer states as predicted by the Born rule. When a small additional dynamics is added, we see the emergence of jumps in the fast collapse regime. The statistics of these jumps can be computed in full generality. Another feature, spikes, seem ubiquitous. However, they are quantitatively understood only in the qubit case.

A first possible generalization is to go from a continuous collapse model to a discrete one, and replace the diffusive equations we had with jump ones. Note that there the jumps we would see are not the same as the emerging jumps between pointers, but could be far smaller jumps in Hilbert space formally equivalent to weak measurements as in the Ghirardi-Rimini-Weber model \cite{ghirardi1986}. So long as the collapses are not exactly projective, they introduces a new timescale, just like the $\gamma$ we had in the continuous case. When the frequency of discrete collapses is sent to infinity, one obtains jumps between pointers that can be quantified exactly just as before \cite{tilloy2016phd,ballesteros2019}. However, while spikes are numerically present as well in this context and seem to have the exact same power law statistics, no proof is known even for the qubit case.

A second generalization would be to characterize spikes precisely for Hilbert spaces of arbitrary finite dimension. This is a surprisingly non trivial task. One reason is that in higher dimensions, it is unclear to know on which submanifold of the Hilbert space the spikes happen, as there is no longer a single path connecting pointers. In general, knowing on which submanifold the trajectories stay for a given collapse operator is a hard question, which has been solved only in simple cases \cite{sarlette2017}.

Finally, it would be interesting to rigorously extend the results we presented here to more realistic situations with Hilbert spaces of infinite dimension. A lot of progress was made on the mathematical physics front in the recent years, by Ballesteros \textit{et al.} \cite{ballesteros2017} in the pure collapse setup and Bauer \textit{et al.} \cite{bauer2018continuous} in the case where collapse competes with other simple dynamics\footnote{An example of such competition in the continuous case is given by the QMUPL model for a free particle where $H\propto \hat{P}^2$ and $\mathcal{O}\propto\hat{X}$. For this particular model, a lot is known rigorously (see \eg \cite{bassi2005,bassi2010} and references therein) and the dynamics is very rich, with behavior already qualitatively distinct from the simpler discrete setting we considered.}. However, the general case and an extension to the even larger Hilbert spaces of quantum field theories remain an open-problem.  


\section{Summary and conclusion}
Continuous collapse models on finite dimensional Hilbert spaces give rise to rich dynamics, already in the simplest case yielding the stochastic differential equation
\begin{equation}\label{eq:simple}
    \upd p_t = \lambda \,(p_\text{eq} - p_t) \,\upd t + \sqrt{\gamma}\, p_t (1-p_t) \,\upd W_t.
\end{equation}
When pure collapse is considered [$\lambda=0$ in \eqref{eq:simple}], one obtains a progressive reduction of the quantum state to one (random) eigenvector (or pointer) of the collapse operator [$p=0$ or $1$ in \eqref{eq:simple}]. This happens because collapse acts as a pure noise term which vanishes only on these eigenvectors, which are thus fixed points of the dynamics. 

The fact that this reduction is progressive and not instantaneous allows to compare its rate $\gamma$ to other system dynamics (unitary or dissipative). Generically, when the collapse process is much faster than other dynamics [$\gamma \gg \lambda$ in \eqref{eq:simple}], we see the (expected) emergence of quantum jumps between the pointers. The latter are not strictly instantaneous, and take roughly $\gamma^{-1}$ to complete. These jumps can be characterized precisely and we note two important facts:
\begin{enumerate}
    \item The jump rates decrease as a function of the collapse rate $\gamma$ when their origin is a Hamiltonian coupling (Zeno effect), whereas they do not depend on $\gamma$ for dissipative transitions (no Zeno effect), as is the case for \eqref{eq:simple},
    \item The jump rates generically depend on the eigenvalues of the collapse operator and not just on the eigenvectors as one would expect for projective measurements.
\end{enumerate}
Finally, the jumps are not the whole story. Perhaps surprisingly, they come decorated with spikes, thin excursions that never complete into jumps. Spikes are power law distributed and prevent any strong form of convergence towards the jump process. They seem ubiquitous and the proof of their existence and rigorous characterization has been done in simple cases. However, their ontological status (are spikes real or just in our mind?) is subtle, especially when collapse models are compared with hidden Markov models.

In the end, it is not entirely clear if the fine characterization of collapse models in the overly simple context we have discussed has any physical relevance. It is possible that spikes, for example, will remain a mathematical curiosity and nothing more. Nonetheless, even then, this study will have put into light a stochastic differential equation \eqref{eq:simple} with a surprisingly rich behavior, that sparked interest in mathematics \cite{bernardin2018,kolb2019}, and even finance \cite{henkel2017}.

\bibliographystyle{apsrev4-1}
\bibliography{main}

\end{document}